\documentclass[sigconf]{acmart}
\acmConference[MSR 2026]{MSR '26: Proceedings of the 23rd International Conference on Mining Software Repositories}{April 2026}{Rio de Janeiro, Brazil}

\usepackage{multirow}
\usepackage{makecell}
\usepackage{tcolorbox}
\usepackage[utf8]{inputenc}
\usepackage{listings}

\AtBeginDocument{%
  }

\copyrightyear{2026}
\acmYear{2026}
\setcopyright{cc}
\setcctype{by}
\acmConference[MSR '26]{23rd International Conference on Mining Software Repositories}{April 13--14, 2026}{Rio de Janeiro, Brazil}
\acmBooktitle{23rd International Conference on Mining Software Repositories (MSR '26), April 13--14, 2026, Rio de Janeiro, Brazil}

\acmDOI{10.1145/3793302.3793580}
\acmISBN{979-8-4007-2474-9/2026/04}

\begin{document}

\title{When is Generated Code Difficult to  Comprehend?\\
Assessing AI Agent Python Code Proficiency in the Wild 
}

\author{Nanthit Temkulkiat}
\authornote{Both authors contributed equally to this research.}
\author{Chaiyong Ragkhitwetsagul}
\authornotemark[1]
\author{Morakot Choetkiertikul}
\affiliation{%
  \institution{Faculty of ICT, Mahidol University}
  \city{Salaya}
  \state{Nakhon Pathom}
  \country{Thailand}
}

\author{Ruksit Rojpaisarnkit}
\affiliation{%
  \institution{Nara Institute of Science and Technology}
  \city{Nara}
  \country{Japan}}

\author{Raula Gaikovina Kula}
\affiliation{%
  \institution{The University of Osaka}
  \city{Osaka}
  \country{Japan}
}

\renewcommand{\shortauthors}{Temkulkiat et al.}

\begin{abstract}
The rapid adoption of AI coding agents is fundamentally shifting software developers' roles from code authors to code reviewers. While developers spend a significant portion of their time reading and comprehending code, the linguistic proficiency and complexity of the Python code generated by these agents remain largely unexplored. This study investigates the code proficiency of AI agents to determine the skill level required for developers to maintain their code. Leveraging the AIDev dataset, we mined 591 pull requests containing 5,027 Python files generated by three distinct AI agents and employed pycefr, a static analysis tool that maps Python constructs to six proficiency levels, ranging from A1 (Basic) to C2 (Mastery), to analyze the code. Our results reveal that: AI agents predominantly generate Basic-level code, with over 90\% of constructs falling into the A1 and A2 categories, and less than 1\% classified as Mastery (C2); AI agents' and humans' pull requests share a broadly similar proficiency profile; High-proficiency code by AI agents are from feature addition and bug fixing tasks. These findings suggest that while AI-generated code is generally accessible to developers with basic Python skills, specific tasks may require advanced proficiency to review and maintain complex, agent-generated constructs.
\end{abstract}



\received{20 February 2007}
\received[revised]{12 March 2009}
\received[accepted]{5 June 2009}

\maketitle

\section{Introduction}

AI coding agents are reshaping software development. With the rise of autonomous AI coding agents such as GitHub Copilot and Cursor, developers and AI agents are mutually contributing code to software projects. Recent empirical findings~\cite{li2025aidev} report that OpenAI's Codex generated over 400,000 pull requests (PRs) across open-source GitHub repositories within less than two months of its release in May 2025.
GitHub also reports that nearly 80\% of new developers use GitHub Copilot in their first week~\cite{octoverse2025}.

The role of software developers is dramatically shifting. 
With the help of AI coding agents, developers not only review code written by humans but also AI-agents~\cite{Takerngsaksiri2025,Takerngsaksiri2025b}.
Reading and comprehending code needs a lot of time and mental energy. A study by Minelli et al.~\cite{Minelli2015} shows that developers spend 70\% of their time reading and understanding code.
Given the breakneck pace of code generation by AI coding agents, the time required to understand code may increase, and reading and understanding AI-generated code has become a key skill for today's software developers.

Due to the steep increase in AI-enabled projects, the number of software projects written in Python is also sharply increasing~\cite{octoverse2025}. 
Python language offers many ways to write code, and a developer's proficiency is essential for using specific language idioms, i.e., \textit{Pythonic} code~\cite{Alexandru2018}. 
Python proficiency is an essential factor in open-source software projects, where contributors need to understand the existing code, especially for newcomers.

\begin{figure}[tb]
    \centering
    \includegraphics[width=\linewidth]{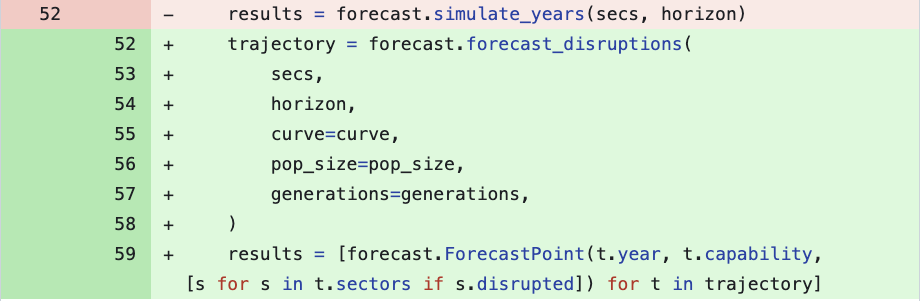}
    \Description[AI agent code snippet]{A code change made by OpenAI Codex using a list comprehension with an \texttt{if} statement (line 59)}
    \caption{A code change made by OpenAI Codex using a list comprehension with an \texttt{if} statement (line 59)}
    \label{fig:motivating_example}
\end{figure}

Nonetheless, a code change made by an AI agent may contain complex code constructs that require Python proficiency to understand correctly.
Figure~\ref{fig:motivating_example} shows a code change in \textsf{AGI-Alpha-Agent-v0} project\footnote{https://github.com/MontrealAI/AGI-Alpha-Agent-v0/pull/774} that an AI agent committed to add a new feature. Line 59 of the added code contains a nested list comprehension with an \texttt{if} statement. Although this code is compact and concise, it is more difficult to understand than its \texttt{for} loop counterpart. Thus, maintaining this code needs a highly proficient contributor. 

\begin{figure*}[tb]
    \centering
    \Description[Study overview]{The overview of this study}
    \includegraphics[width=0.8\linewidth]{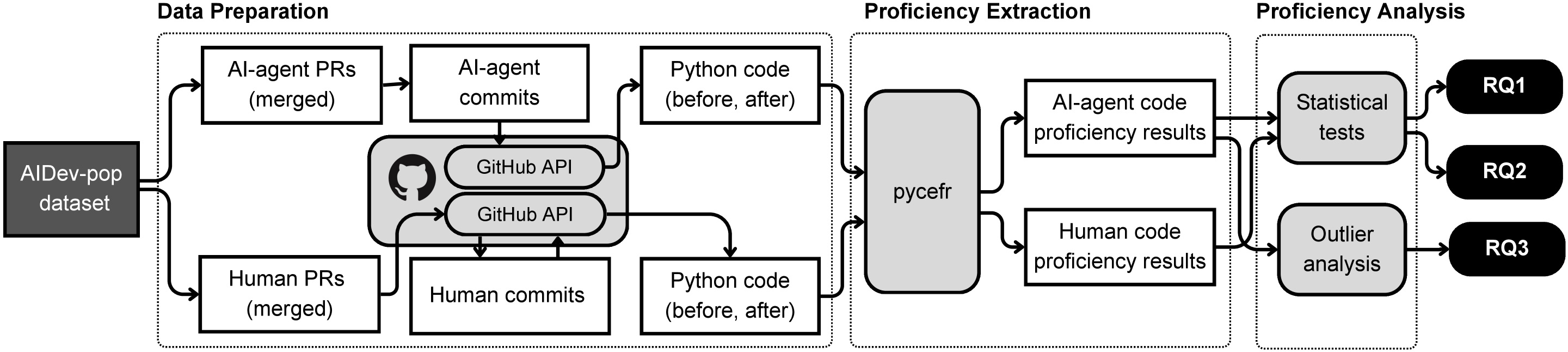}
    \caption{The overview of this study}
    \label{fig:overview}
\end{figure*}

While prior research has explored how external factors like prompt English proficiency influence the level of AI-generated code~\cite{Rojpaisarnkit2025}, there is currently a lack of research directly comparing the proficiency of AI-contributed code against human-written code in real-world software projects. Thus, this study fills in the gap. Leveraging the large-scale Python code in PRs created by AI coding agents from the AIDev dataset~\cite{li2025aidev}, we analyzed and extracted the coding proficiency of their contributed code constructs using an automated tool for Python proficiency analysis called pycefr~\cite{Robles2022}. 
We aim to answer the following research questions in this study.

\begin{itemize}
\item RQ1: \textit{What are the Python proficiency levels of AI agents' code?} 

\item RQ2: \textit{What are the differences between the Python proficiency levels of AI agents' and humans' written code?} 

\item RQ3: \textit{What kinds of AI agents' pull requests contain the most proficient code?} 
\end{itemize}

To the best of our knowledge, this is the first study to compare the coding proficiency of AI coding agents with human-written code and paves the way for future studies in this area. A replication package is made publicly available\footnote{Replication package: \url{https://github.com/EarnGH/AIDevMSR2026}}.

\section{Background} 
\subsection{CEFR and Python Code Proficiency Analysis} 
\label{sec:cefr} 
The Common European Framework of Reference for Languages (CEFR)~\cite{cefr} constitutes an internationally standardized framework for assessing linguistic proficiency. It classifies language skills into six levels: A1 (breakthrough) and A2 (waystage) for basic users; B1 (threshold) and B2 (vantage) for independent users; and C1 (advanced) and C2 (mastery) for proficient users. By providing this unified metric, the framework streamlines the evaluation of competence for educators and learners alike. Furthermore, the CEFR enables the alignment of national qualifications and facilitates skill verification for academic institutions and employers.

\begin{table}[tb]
    \caption{Python proficiency levels in pycefr}
    \label{tab:cefr}
    \begin{tabular}{cp{5cm}}
        \toprule
        \textbf{Level group} & \textbf{Level and construct examples} \\ 
        \midrule
        \multirow{2}{*}{\makecell{A\\Basic user}} & A1: Breakthrough or beginner (e.g., \texttt{print}, \texttt{if} statement, list) \\  
         & A2: Waystage or elementary (e.g., \texttt{open} function (files), nested list) \\ 
         \midrule
        \multirow{2}{*}{\makecell{B\\Independent user}} & B1: Threshold or intermediate (e.g., list with dictionary, \texttt{with}) \\ 
        & B2: Vantage or upper intermediate (e.g., list comprehension, \_\_dict\_\_attribute) \\ \midrule
        \multirow{2}{*}{\makecell{C\\ Proficient user}} & C1: Effective operational proficiency or advanced (e.g., \_\_slots\_\_, generator) \\ 
        & C2: Mastery or proficiency (e.g., meta-class, decorator class) \\
        \bottomrule
    \end{tabular}
\end{table}

Inspired by CEFR, pycefr~\cite{Robles2022} is a static analysis tool that automates the assessment of code proficiency in Python projects. 
Given Python code, pycefr identifies the code constructs used and estimates the skill level required to understand them.
By adapting the CEFR's linguistic levels to the domain of programming, the tool maps Python code constructs to six distinct levels of complexity (see Table~\ref{tab:cefr}). For instance, fundamental constructs such as \texttt{if} statements and nested lists are categorized as Basic (A1 and A2, respectively). \texttt{break} statements and list comprehensions are mapped to the Intermediate levels (B1 and B2), while advanced features, including generator functions and metaclasses, correspond to the Proficient levels (C1 and C2). The complete list of code constructs and their CEFR levels can be found in the pycefr repository~\cite{pycefr}.
It is used in a recent study to analyze the progression of proficiency in Python OSS projects~\cite{Charatvaraphan2025}.
In this study, we employ pycefr to analyze the proficiency of AI agents' code and compare it with that of humans.

\section{Study Design}
The overview of the study is depicted in Figure~\ref{fig:overview}. 
We use the AIDev dataset~\cite{li2025aidev}, a large-scale dataset of AI agent PRs, and select the AIDev-pop subset, which includes only repositories with more than 100 stars, focusing on popular repositories (filtered by stars), and excludes most toy projects.

\subsection{Data Preparation}
In this study, we focus exclusively on Python repositories and analyze code changes extracted from PR-level commits for both AI agents and human developers. 
Thus, we first extracted the PR, repository, and commit-related information required for code-level analysis. We then associated each PR with its corresponding project and programming language, and further merged commit-level information to obtain detailed file-level changes linked to each PR.
After filtering for only Python projects, we filtered for merged PRs, as they represent the code changes that were accepted. Lastly, for the commits, we filtered only the commits in which AI coding agents were the authors (e.g., \texttt{devin-ai-integration[bot]}). Nonetheless, Claude Code and OpenAI Codex attribute their contributions to human developers~\cite{li2025aidev}. Thus, the authors of the commits using them were varied, and we could not precisely decide which commits were actually made by agents. To ensure the validity of our analysis, we restricted ourselves to analyzing only the three remaining agents: Copilot, Cursor, and Devin. 

\begin{table}[tb]
    \centering
    \caption{Summary of the collected and analyzed data for RQ1}
    \label{tab:metrics}
    \begin{tabular}{lrrrr}
        \toprule
        Category & PRs & Repos & Commits & Files \\
        \midrule
        Collected merged PRs & 591 & 145 & 1,830 & 5,027 \\
        Analyzed by pycefr & 524 & 139 & 1,463 & 3,441 \\
        \bottomrule
    \end{tabular}
\end{table}

As shown in Table~\ref{tab:metrics} (Collected merged PRs section), this resulted in 591 PRs from 145 Python projects, containing 1,830 commits and associated with 5,027 Python files.
Next, we retrieved the added code in each commit. pycefr only works on complete Python files. Thus, for each Python file, we extracted the patch information from the commit details (i.e., line-level diffs containing code additions and deletions) and queried the GitHub API to retrieve the file's version before the commit to apply the diffs.
Using this information, we reconstructed both the \emph{before} and \emph{after} versions of each file. Files introduced for the first time in a PR were identified as newly created files and were represented by a single (after) version.
For human code, we filtered the merged PRs and queried the GitHub API to obtain the associated commits and the code changes. Finally, we retrieved 785 PRs from 384 Python repositories, containing 3,548 commits and covering 20,063 Python source code files.

\subsection{Proficiency Extraction}
We applied pycefr to Python source code files, both agent-generated and human-written.
pycefr failed to analyze some files due to parsing issues with the abstract syntax tree. Thus, the total number of Python code files analyzed successfully by pycefr was slightly lower than the ones we initially retrieved. As displayed in Table~\ref{tab:metrics} (Analyzed by pycefr section), for AI agents, there were 3,441 Python files associated with 1,463 commits across 524 PRs in 139 repositories. For humans, there were 12,366 Python files associated with 2,814 commits across 694 PRs in 151 repositories successfully analyzed.

For newly added files, all detected constructs and their associated proficiency levels were counted directly. For modified files with both before and after versions, we performed a differential analysis by matching constructs between the two versions and computing the difference in their occurrence counts. Only constructs newly introduced in the later version were retained, while negative differences were clipped at zero, similar to the previous study~\cite{Charatvaraphan2025}.
This procedure ensures that the proficiency analysis reflects only the code introduced in a given PR, rather than the entire file content, which may include contributions from prior commits authored by human developers or other agents. Moreover, we focus only on the added code and ignore the deleted code by AI agents, as that is what the human contributors need to maintain. %

\subsection{Proficiency Analysis}
\paragraph{Methodology to answer RQ1} We aggregated proficiency results, i.e., the number of code constructs classified into one of the six Python proficiency levels, across all AI coding agents to examine their distribution. Then, we divided the proficiency scores of each AI agent and compared the differences among them.

\paragraph{Methodology to answer RQ2} We filtered the repositories to include only those in which both AI agents and human programmers contributed code. Then, we compared the distributions of code constructs across Python proficiency levels in AI-generated and human-written code.

\paragraph{Methodology to answer RQ3} We identified PRs in which AI agents introduce unusually high amounts of proficient-level Python constructs. We aggregated proficiency scores at the PR level by summing the counts of C1 and C2 constructs introduced by AI agents across all files and commits in each PR. We then detected PRs with exceptionally large C1 + C2 totals using the IQR-based outlier criterion ($> Q_3 + 1.5\times IQR$). 
Using these outliers, we extracted 11 task categories (e.g., \texttt{feat}, \texttt{fix}, \texttt{docs}) associated with the PRs from the AIDev dataset. 

\section{Results}

\begin{table*}[tb]
\centering
\caption{Distribution of AI agents' Python code constructs classified into proficiency levels. The highest values are highlighted.}
\label{tab:ai-proficiency-by-agent}
\begin{tabular}{l r r r r r r}
\toprule
Agent & A1 & A2 & B1 & B2 & C1 & C2 \\
\midrule
All agents
& 91,223 (35.72\%)
& 138,894 (54.39\%)
& 12,639 (4.95\%)
& 8,913 (3.49\%)
& 2,731 (1.07\%)
& 972 (0.38\%) \\
\midrule
Copilot
& \textbf{51,807 (36.85\%)}
& 75,857 (53.96\%)
& 6,614 (4.70\%)
& 4,080 (2.90\%)
& \textbf{1,655 (1.18\%)}
& 561 (0.40\%) \\
Cursor
& 15,959 (31.76\%)
& \textbf{29,449 (58.61\%)}
& 2,337 (4.65\%)
& 1,755 (3.49\%)
& 521 (1.04\%)
& \textbf{221 (0.44\%)} \\
Devin
& 23,467 (36.35\%)
& 33,588 (52.02\%)
& \textbf{3,688 (5.71\%)}
& \textbf{3,078 (4.77\%)}
& 555 (0.86\%)
& 190 (0.29\%) \\
\bottomrule
\end{tabular}
\end{table*}

\subsection{RQ1: What are the Python proficiency levels of AI agents’ code?}

\subsubsection{Overall Proficiency of AI Coding Agents}
Table~\ref{tab:ai-proficiency-by-agent} (first row) summarizes the number of Python code constructs by the CEFR-based proficiency levels for all the agents. The results show that the majority of AI-generated Python code constructs fall within the Basic proficiency range. In particular, A2 (Waystage) is the most dominant level, accounting for 54.39\% of all code constructs, followed by A1 (Breakthrough) (35.723\%). Together, Basic-level constructs (A1 and A2) comprise more than 90\% of the AI agents’ contributed code.
Intermediate-level constructs appear far less frequently. B1 (Threshold) accounts for 4.95\%, while B2 (Vantage) represents only 3.49\%. Proficient-level constructs are relatively rare: C1 (Advanced) and C2 (Mastery) contribute 1.07\% and less than 0.5\%.

\subsubsection{Proficiency of Each AI Coding Agent}
Table~\ref{tab:ai-proficiency-by-agent} presents the proportion of CEFR levels for code generated by Copilot, Cursor, and Devin.
We can clearly see that Copilot is the most widely adopted agent among the three agents, with the number of analyzed code constructs far exceeding those of the other agents.
Across all agents, Basic-level constructs (A1 and A2) dominate the generated code, similar to the overall result. For all three agents, A2 is the most prevalent (52.02\% to 58.61\%). A1 is the second most common level (31.76\% to 36.85\%).
Intermediate-level constructs (B1 and B2) appear less frequently but consistently across agents. B1 ranges from 4.65\% to 5.71\%, while B2 contributes between 2.90\% and 4.77\%. %
Proficient-level constructs are relatively rare for all models. C1 remains below 1.20\%, while C2 accounts for less than 1\% of the generated code in all cases. Copilot and Cursor show the highest proportion of C1 and C2 constructs (1.18\% and 0.44\%), respectively.

A chi-square test of independence indicates that proficiency distributions differ statistically across AI agents ($\chi^2 = 1,117.56$, $df = 10$, $p < 0.001$); however, the effect size is small (Cramér’s $V = 0.046$), suggesting that the observed differences are minor in magnitude. A rank-based Kruskal–Wallis test yields consistent results ($H = 343.41$, $p < 0.001$).

\textbf{Answer to RQ1:} AI coding agents predominantly generate Python code that relies on beginner and elementary language constructs, with fewer contributions involving intermediate and proficient Python features.
There are statistically significant differences in proficiency distributions across agents with a small effect size.

\subsection{RQ2: What are the differences between the Python proficiency levels of AI agents' and humans' written code?}

After filtering the AIDev dataset for repositories in which both AI agents and human programmers contributed code, we obtained 39 repositories associated with 230 PRs, 647 commits, and 1,527 files for agents and 385 PRs, 1,348 commits, and 6,950 files for humans.

As shown in Table~\ref{tab:ai-vs-human-same-projects}, for both AI agents and humans, Basic-level constructs (A1 and A2) dominate the generated code. AI-generated code consists primarily of A2 and A1 constructs, while human-written code exhibits a similar pattern. Together, Basic-level constructs comprise more than 88\% of the code in both groups. The Intermediate-level constructs (B1 and B2) appear at comparable frequencies in AI-generated and human-written code. In contrast, Proficient-level constructs (C1 and C2) remain rare for both AI and humans, accounting for less than 2\% of total occurrences in each case. Nonetheless, the AI agents generated slightly more C1 code than humans (1.12\% vs.~0.85\%). On the other hand, human programmers created slightly more C2 code than AI agents (0.57\% vs.~0.42\%).
A chi-square test indicates a statistically significant difference ($\chi^2 = 1,124.52$, $df = 5$, $p < 0.001$) with a small effect size (Cramér’s $V = 0.041$), similar to the rank-based Kruskal–Wallis test ($H = 265.59$, $p < 0.001$).

\textbf{Answer to RQ2:} AI-generated and human-written code exhibit statistically significant differences in proficiency distributions. Nonetheless, their overall proficiency profiles are broadly 
similar. 
Agents generated slightly more C1 code than humans.

\begin{table}[tb]
\caption{Proficiency distribution for agents and human code}
\label{tab:ai-vs-human-same-projects}
\begin{tabular}{l r r r r r r}
\toprule
 & A1 & A2 & B1 & B2 & C1 & C2 \\
\midrule
AI agents & 35.52\% & 52.85\% & 5.14\% & 4.94\% & 1.12\% & 0.42\% \\
Human & 31.60\% & 57.70\% & 5.37\% & 3.91\% & 0.85\% & 0.57\% \\
\bottomrule
\end{tabular}
\end{table}

\begin{figure}[tb]
    \centering
    \Description[Outlier agent PR tasks]{Tasks of the outlier agent PRs with a high number of proficient code (C1 + C2) (log scale)}
    \includegraphics[width=0.9\linewidth]{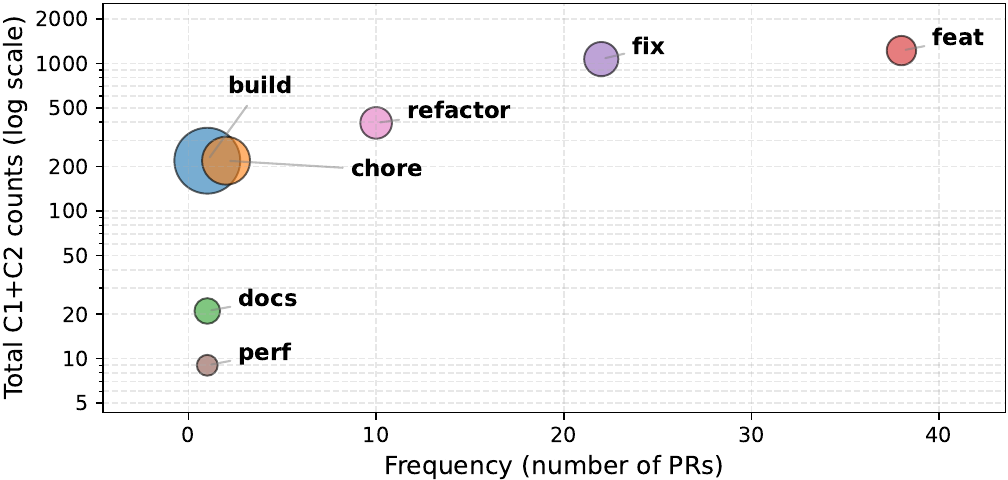}
    \caption{Tasks of the outlier agent PRs with a high number of proficient code (C1 + C2) (log scale)}
    \label{fig:action_impact}
\end{figure}

\subsection{RQ3: What kinds of AI agents' pull requests contain the most proficient code?}

In total, we identify 79 outlier agent PRs with the C1 + C2 counts ranging from 8 to 339, with a median of 19 and an average of 41.41.
After analyzing these PRs, we identified seven major tasks. 
The scatter plot in Figure~\ref{fig:action_impact} depicts the seven tasks with their number of PRs and total counts of C1 + C2 (in the log scale). The size of the bubbles represents the average count of C1 + C2
of each task.
The most frequent task is feature development (\textit{feat}) (38 PRs with 1,220 C1 + C2 code constructs), followed by bug remediation (\textit{fix}) (22 PRs, 1,069 constructs), \textit{refactor} (10 PRs, 395 constructs), 
\textit{chore} (2 PRs, 219 constructs), \textit{build} (1 PR, 219 constructs), \textit{docs} (1 PR, 21 constructs), and performance improvement (\textit{perf}) (1 PR, 9 constructs). Interestingly, the task with the highest average of C1 + C2 is \textit{build} (219 per PR), \textit{chore} (109.50), and \textit{fix} (48.59).

\textbf{Answer to RQ3:} PRs that perform new feature addition contain the highest total number of proficient code. However, PR that relate to building the project contain the highest average number of proficient code per PR.

\section{Discussion and Future Work}
This study finds that all AI coding agents exhibit a similar distribution of Python code proficiency, with slight differences among the agents and comparable to that of human programmers. This implies that maintaining agents' code may not differ from that of human counterparts. Nonetheless, some of the agents' PRs contain highly proficient code, especially in adding features and fixing bugs. Thus, maintainers need to pay attention to such PRs when assigning reviewers, ensuring the introduced code is reviewed correctly.

Future work includes (1) a deeper investigation of the scenarios and reviews of agents' PRs, and determining whether the high proficiency affected the review process; (2) analyzing proficiency in AI agents' vs. humans' code per task (e.g., feature addition, refactoring) across project types (e.g., AI vs. cryptocurrency); (3) studying standardized agent prompting languages, which might be necessary to ensure consistent code proficiency across an organization; (4) refactoring agents' and humans' code based on proficiency.

\bibliographystyle{ACM-Reference-Format}
\bibliography{references}

\end{document}